\documentclass[]{aastex631}

\usepackage{color}

\shorttitle{AASTeX v6.3.1 Sample article}
\usepackage{float}
\usepackage{booktabs}
\usepackage{subfigure}
\usepackage{graphicx}
\usepackage{footnote}
\usepackage{multirow}
\usepackage{color}
\graphicspath{{./}{figures/}}

\begin{document}

\title{Global Structure of Accretion Flows in Sgr A*}

\author[0009-0002-0726-9005]{Shenyue Yin}
\affiliation{School of Physical Science and Technology, Southwest Jiaotong University, Chengdu 610031, People's Republic of China}

\author[0000-0003-1039-9521]{Siming Liu}
\affiliation{School of Physical Science and Technology, Southwest Jiaotong University, Chengdu 610031, People's Republic of China}

\email{liusm@swjtu.edu.cn (SML); qyhy@my.swjtu.edu.cn}

\begin{abstract}
Sagittarius A* (Sgr A*) is a compact radio source at the Galactic center. Observations have confirmed that its mass is approximately (4.1±0.4)×10$^{6}$ M$_{\odot}$, and Sgr A* is generally believed to be powered by gas accretion onto a supermassive black hole. Multifrequency radio observations of the pulsar J1745-2900, about 0.12 pc away from Sgr A*, reveal an unusually large Faraday rotation. Combined with X-ray observations, this indicates that there is a strong magnetic field (greater than 8 mG) leading to a low $\beta$ plasma at large scales.
We show that the gas starts to be captured by the black hole below tens of thousands of the Schwarzschild radii $r_S$, where the gas pressure starts to dominate. Assuming that the accretion rate along magnetic fields at large scales decreases with the distance to the black hole following a power law, it is shown that, with an accretion disk below tens of $r_S$, as revealed with the EHT observations, 
there should be a supersonic wind above such a small accretion disk, and the accretion flow may be convection-dominated from tens of $r_S$ to tens of thousands of $r_S$. Detailed modeling is warranted.
\end{abstract}


\section{Introduction} \label{sec:intro}

Sgr A* is a compact, non-thermal radio source located in the center of the Milky Way \citep{1974ApJ...194..265B}, approximately 8.4 kpc from the Earth. Observations of OB stars \citep{2003ApJ...586L.127G, 2008ApJ...689.1044G, 2009ApJ...707L.114G, 2009ApJ...692.1075G, 2012Sci...338...84M} show that the mass of Sgr A*  is (4.1±0.4)×10$^{6}$ M$_{\odot}$. The IR flux upper limit plus an X-ray detection with Chandra \citep{2001Natur.413...45B, 2003ApJ...591..891B} indicate that Sgr A*'s bolometric luminosity is extremely low:\textit{  }$L_{bol}$$\approx$10$^{36}$ ergs s$^{-1}$ $\approx$ 3×10$^{-9}$\textit{ }$L_{\rm Edd}$. If the radiative efficiency of the Sgr A * accretion flow is $\epsilon \simeq 10\%$, the accretion rate $\dot{M}$ of Sgr A* must be about 10$^{17}$ g s$^{-1}$, which is compatible to the upper limit of approximately 2×10$^{-7}$ M$_{\odot}$ yr$^{-1}$ inferred from millimeter, IR and X-ray observations \citep{1999ApJ...517L.101Q, 2007ApJ...654L..57M}.

In 2013, the NASA's Swift X-ray Telescope detected a bright X-ray flare 0.12 pc away from Sgr A* \citep{2013ApJ...770L..24K, 2013ApJ...775L..34R}, and the NASA's NuSTAR telescope subsequently reported a pulsation period of 3.76 s at that location \citep{2013ApJ...770L..23M}, discovering a magnetar in outburst, namely PSR J1745-2900. Radio emission from PSR J1745-2900 exhibits high linear polarization \citep{2013MNRAS.435L..29S}. The rotation measure (RM) to this pulsar was determined to be (-6.696±0.005)×10$^{4}$ rad m$^{-2}$ using a RM synthesis method \citep{2013Natur.501..391E, 2018ApJ...852L..12D}, which is higher than any Galactic sources other than Sgr A*. The gas density profile n(R) is estimated to be $\approx$ 26 cm$^{-3}$(R/0.4 pc)$^{-1}$, where\textit{ R} is the distance from Sgr A* \citep{2003ApJ...591..891B, 2013Sci...341..981W}. Since RM = 8.1×10$^{5}$[B$_{||}(R)$/G][n(R)/cm$^{-3}$][R/pc] rad m$^{-2}$, where B$_{||}$(r) is the magnetic field along the line of sight, at R=0.12 pc, one finds a magnetic field B $\gtrsim$ 8 [RM/($6.7\times 10^4$ rad m$^{-2}$)][n\textit{$_{0}$}/26 cm$^{-3}$]$^{-1}$ mG.

\textbf{Recent polarimetric observations of Sgr A* at 230 GHz by the Event Horizon Telescope (EHT) provide new strong constraints. The EHT 2017 observations measured the Faraday rotation at this frequency$\sim$-4.6 × 10$^{5}$ rad m$^{-2}$ \citep{2024ApJ...964L..25E, 2007ApJ...654L..57M}, corresponding to an electric vector polarization angle rotation of approximately -46°. This value is significantly higher than PSR J1745-2900, indicating extremely strong magnetization of plasma near the black hole's event horizon. It is important to note that the RM measured by the EHT incorporates contributions from both a potential external Faraday screen and an internal emission region. Its rapid temporal variation (50$\%$ on subhour timescales)  suggests significant Faraday rotation in the internal emission region, while the long-term stable sign (decades-long minus sign) points to a slowly evolving external screen. In this work, we interpret the EHT RM observations as evidence for the persistent existence of strong magnetic fields (B $\gtrsim$ 8 mG) from tens to tens of thousands Schwarzschild radii ($r_S$).}

The X-ray emitting gas pressure at 0.12 pc is $\sim$10$^{-8}$ erg cm$^{-3}$, which is much lower than the magnetic pressure $\sim$10$^{-6}$  erg cm$^{-3}$. Therefore, gases must move along magnetic field lines at large distances from Sgr A* in the absence other pressures. \textbf{\citet{2019MNRAS.482L.123R} carried out detailed MHD simulations of stellar winds around Sgr A* and attributed the high RM to collisions of strong magnetized winds. They predicts variability of these RMs on a timescale of hundreds of years and significant variation of the RM over a scale of $\sim 0.01$ pc, the characteristic size of the wind shocks. We assume here the presence of a strong large scale magnetic field. The RM should follow the distribution of the column density of free electrons and should be relatively stable. }

{\bf The basic setup of this work is quite different from \citet{1974Ap&SS..28...45B, 1976Ap&SS..42..401B} (hereafter BKR74 and BKR76) and \citet{2002ApJ...566..137I}, where the magnetic pressure is negligible at large distances from the center object. BKR74 investigated the effects of spherical accretion of matter onto collapsing stars on a dynamically trivial frozen-in magnetic field, and found that magnetic fields undergo significant amplification during the accretion process at small radii. 
BKR76 considered effects of the enhanced magnetic field on the accretion flow and constructed an axis-symmetric accretion model. In BKR76, the authors divided the flow region into three zones: }

\textbf{(i) The subsonic region, located far from the star, where the magnetic field is dynamically unimportant, and the flow is nonstationary and nearly spherically symmetric. In this region the magnetic energy density never exceeds the kinetic energy density of accreting gases.}

\textbf{(ii) The transition region between the subsonic and supersonic zones, where the flow is stationary, the magnetic field begins to be enhanced, and the flow speed approaches the speed of sound.}

\textbf{(iii) Near the star, a stationary supersonic flow emerges where the magnetic energy density and the kinetic energy density are in equipartition. A stationary disk forms in the equatorial plane($\theta=\pi/2$), the structure of the disk is determined by magnetic dissipation mechanisms.}

\textbf{Based on the two aforementioned papers, \citet{2003PASJ...55L..69N} proposed a Magentically Arrested Disk (MAD), which assumes that accretion flows accumulate substantial polar magnetic flux inward, and accretion pressure prevents the fields from escaping. Consequently, the accumulated polar magnetic field disrupts the axisymmetric accretion flow at the magnetospheric radius $R_m=r_mr_S$. Beyond $R_m$, the flow remains largely axisymmetric, while within $R_m$, the magnetic field disrupts and fragments it into blobs and streams. At this point, the gas velocity is far below the free-fall velocity, providing an explanation for a slow radial motion into the black hole. Similar mechanism may also suppress accretion at large scales as explored below in this paper. 
}

\textbf{\citet{2002ApJ...566..137I} carried out MHD simulations of the spherical accretion with a weak uniform magnetic field at large scales, and found 
that efficient magnetic reconnection heats the gas as the magnetic field is dragged inward by the accretion flow. As the gas moves outward, it further stretches magnetic field lines, ultimately forming a high-density, low-radial-velocity convective core near the black hole. 
In the three-dimensional MHD simulations of Radiatively Inefficient Accretion Flows (RIAF)\citep{2003ApJ...592.1042I}, the authors considered two distinct magnetic field configurations. When the initial magnetic field is poloidal, the system underwent a two-stage evolution. During the initial transient phase, the gas formed a Keplerian disk accompanied by thermal corona and outflows at the poles. As magnetic flux continued to accumulate near the black hole, the magnetic field eventually disrupted the disk's axial symmetry. This culminated in a system dominated by a strong dipole magnetic field (with plasma $\beta\sim 10^{-3}$ to $10^{-4}$) across most regions, accompanied by several radially narrow accretion flows. When the simulation finally reached a steady state, an MAD-like system formed. 
When the initial magnetic field is toroidal, the simulation uncovered a Convection Dominated Accretion Flows (CDAF)-like structure, with the flow interior dominated by intense convective turbulence driven by entropy gradients resulting from energy release during magnetic reconnection. Magnetic stresses transport angular momentum outward while Reynolds stresses transport angular momentum inward. Convection significantly suppresses radial mass transport. The radial profile of the density is relatively flat, approximately following $\rho\propto R^{-1}$, which is flatter than the standard ADAF ($\rho\propto R^{-3/2}$), but slightly steeper than the CDAF theoretical prediction ($\rho\propto R^{-1/2}$). The authors attributed this to leakage of the convective energy flux in the vertical direction.}

\textbf{\citet{2018MNRAS.478.3544R} simulated how Sgr A* accretes through the stellar winds from its surrounding 30 Wolf-Rayet stars, and proposed the stellar-wind-fed model. The structure of the model at large scales (R$\gtrsim$0.4 pc) is dominated by outflows. In this region, stellar winds collide to form bow shocks, heating the gas to $2 \times 10^8$ K, and the flow follows the standard Parker stellar wind solution with $\rho\propto R^{-2}$, reproducing the diffuse thermal X-ray emission observed with the {\it Chandra}. Only a fraction of low angular momentum gas forms a two-component accretion structure in the inner region (R$\lesssim$0.01 pc) with most accretion occurring in the polar regions. A small fraction of the gas forms a pressure-supported sub-Keplerian disk near the equatorial plane. Within the accretion region, the density distribution follows $\rho \propto R^{-1}$, and accretion rate follows $\dot{M} \propto R^{1/2}$. 
Subsequently, \citet{2020MNRAS.492.3272R} conducted detailed MHD simulations of the stellar wind-fed accretion model with the initial magnetic field determined by a parameter $\beta_w$, the ratio between ram pressure of the wind and its magnetic pressure at the equator and found that even with an extremely weak initial magnetic field($\beta_w =10^6$), the field can be amplified to $\beta < 10$ through flux freezing at R$\sim$10$^{-4}$ pc. However, the magnetic field has few effects on the accretion flow, which remains dominated by low angular momentum gases, similar to the hydrodynamical simulation. The primary role of the magnetic field is to twist high angular momentum gas from the equatorial plane toward the poles via large-scale magnetic torque, thereby forming magnetically driven outflows.}

In this work, we propose a model of accretion flows along magnetic field lines at large scales,  
and find that the gas pressure starts to overcome the magnetic pressure below tens of thousands of the Schwarzschild radii. 
In Section 2, we give the basic equations of the accretion model and analyze the pressure profile.
An accretion model is analyzed and discussed in Section 3. In Section 4, we discuss the model implications and draw conclusions.

\section{Basic Equations For The Accretion Flow}

\begin{figure}[h]
    \centering
    \includegraphics[trim={0 0.cm 0 0}, clip, width=0.4\textwidth]{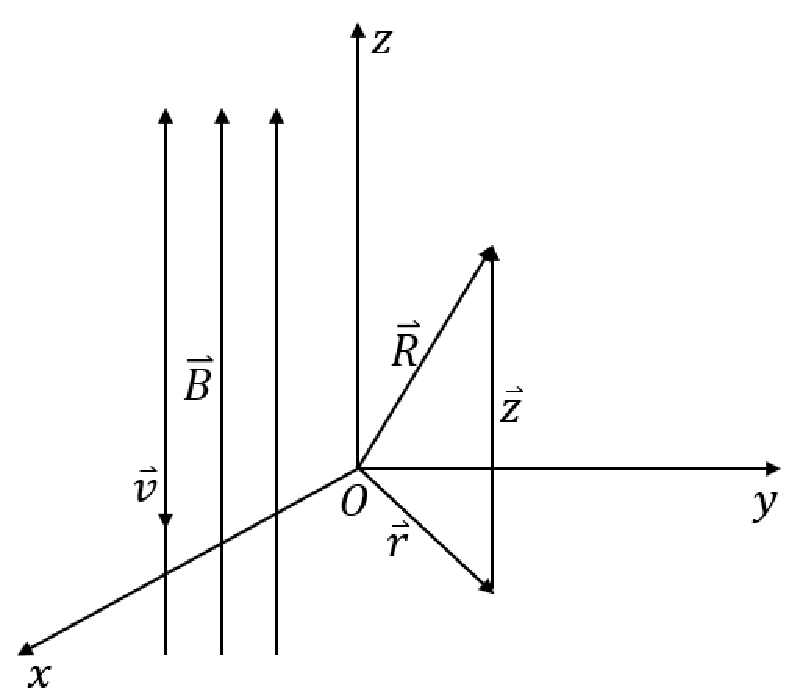}
    \caption{The {\bf cylindrical} coordinates adopted in this study. The magnetic field $\vec{B}$ is in the $z$ direction. 
    \textbf{$\vec{r}$ is the distance of magnetic field line to the origin, where the black hole locates. $\vec{R}$ indicates the location of a gas element, that moves along a magnetic field line with a velocity $\vec v$ in the magnetic pressure dominated region at large radii.}
    }
    \label{fig:1}
\end{figure}

The Bondi accretion model studies spherically symmetric accretion onto a point mass \citep{1952MNRAS.112..195B}. The accretion flow model we propose is similar to the Bondi accretion model to study one dimensional accretion along magnetic field lines. In the model, we assume that the magnetic field is uniform and in perpendicular to the black hole's equatorial plane (see Fig. \ref{fig:1}). The accretion material is modeled as compressible ideal adiatic gases. In reality, the accretion flow may cross magnetic field lines near the equatorial plane due to turbulent dissipation forming an accretion disk, especially in gas pressure dominated regimes. 

\begin{figure}[h]
    \centering
    \includegraphics[trim={0 0.cm 0 0}, clip,width=0.4\textwidth]{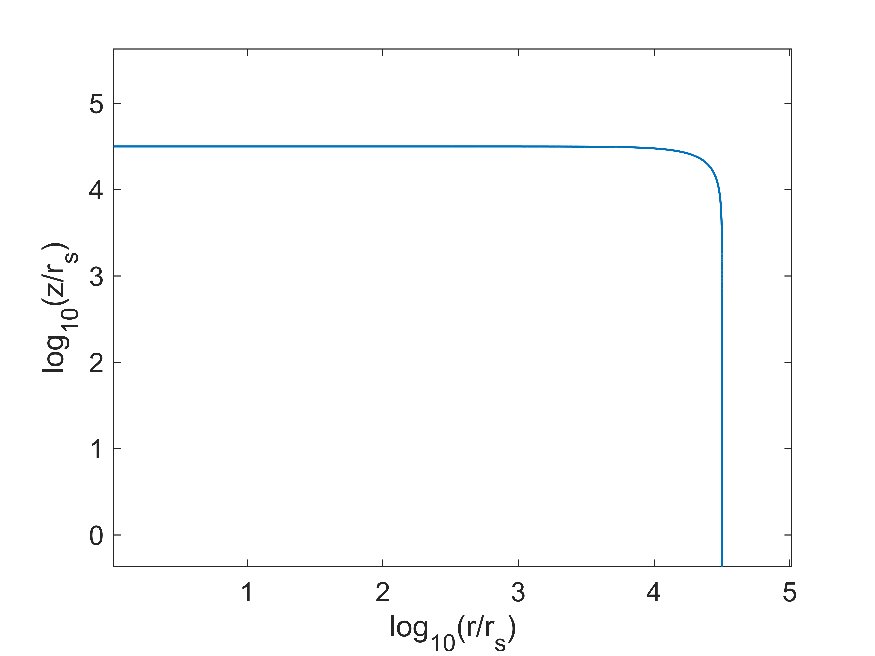}
    \includegraphics[trim={0 0.cm 0 0}, clip,width=0.4\textwidth]{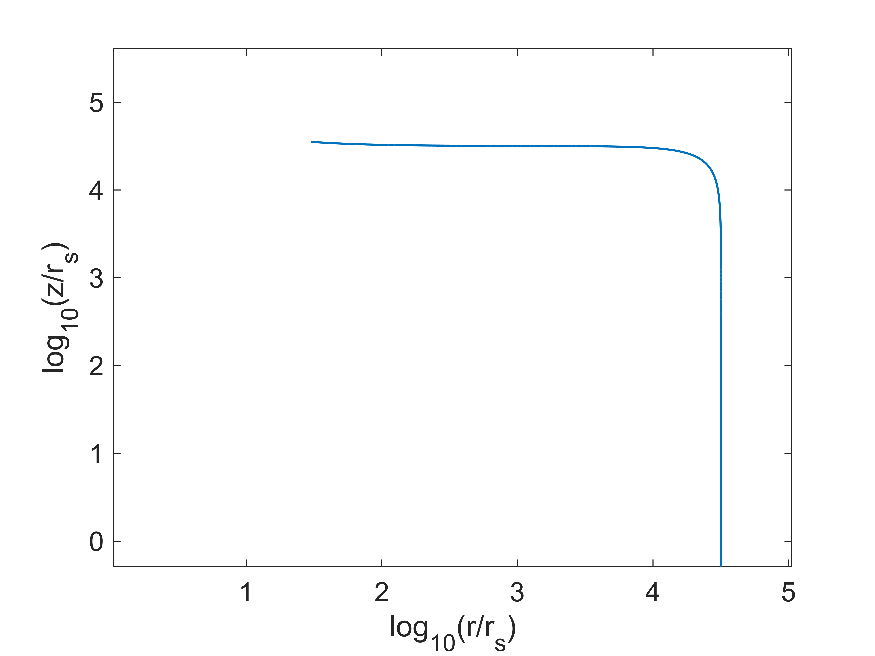}
    \caption{Locations where the gas pressure is equal to the magnetic pressure for $B=8$ mG. \emph{Left}: For the case of pressure equilibrium. \emph{Right}: For an accretion model with $r_0=30\ r_S$, $\alpha=-0.5$.
    }
    \label{fig:2}
\end{figure}

Then we have
\begin{equation}\label{eq:momentum}
v{{\rm d} v \over {\rm d} z} +{{\rm d} p\over \rho {\rm d} z}+{GMz\over(r^2 +z^2)^{3/2}} =0,
\end{equation}
where $v$ is the gas speed, $p$ is the gas pressure, $\rho$ is the mass density, $G$ is the gravitational constant, $M$ is the mass of the central body. The equation of state is given by
\begin{equation}\label{eq:adiabatic}
{p\over p_{0}}=\left({\rho\over \rho_{0}}\right)^\gamma\,,
\end{equation}
where $p_{0}$,  $\rho_{0}$ denote the gas pressure and mass density at the outer boundary, $\gamma$ is the adiabatic index of  gas.
In the case of pressure equilibrium,
\begin{equation}\label{eq:nonaccmomentum}
{{\rm d} p\over \rho{\rm d}z}+{GMz\over(r^2 +z^2)^{3/2}} =0\,.
\end{equation}
By adopting the boundary conditions of $n=26$ cm$^{-3}$, gas temperature $kT =3.5$ keV, where $k$ is the Boltzmann constant, at $R=0.4$ pc with $\gamma=5/3$, the left panel of Figure \ref{fig:2} shows the location where the gas pressure is equal to the magnetic pressure of $5.1\times 10^{-6}$ erg cm$^{-3}$. {\bf Below $R_0\simeq 30000 r_S$, the gas pressure exceeds the magnetic pressure. We note that since no radiative energy loss is considered, the sound speed is comparable to the Keplerian speed and the gas pressure is comparable to the potential energy density. The magnetic field in the gas pressure dominated region will be amplified and stretched due to gravity of the central object, and toroidal currents are expected \citep{1976Ap&SS..42..401B}.}

Following \citet{1999MNRAS.303L...1B}, we assume that the accretion rate along magnetic field lines is given by
\begin{equation}\label{eq:rhov}
\rho v=\rho_0 c_{s} \left({r\over r_0}\right)^\alpha.
\end{equation}
where $c_s$ is the sound speed at the outer boundary, $\alpha$ is a dimensionless parameter that, together with $r_0$, determines the total accretion rate. For a negative value of $\alpha$, this assumption implies that the flow must be supersonic at infinity for $r<r_0$, indicating an outflow solution.
With $r_0=30\ r_S$, where $r_S=1.27\times 10^{12}$ cm is the Schwarzschild radius of black hole, $\alpha=-0.5$, $c_s=5.5\times 10^{7}$ cm s$^{-1}$, $\rho_0 =4.3\times10^{-23}$ g cm$^{-3}$, the right panel of Figure 2 shows that, for $r>r_0$, the pressure profile has a very weak dependence on the accretion rate. These results show that the gas pressure will dominate the magnetic pressure within a few tens of thousands of $r_S$. One therefore may have convection-dominated accretion flows (CDAF) as studied in detail by \citet{2000ApJ...539..798N, 2002ApJ...566..137I}.


\begin{figure}[h]
    \centering
    \includegraphics[trim={0 0.cm 0 0}, clip,width=0.4\textwidth]{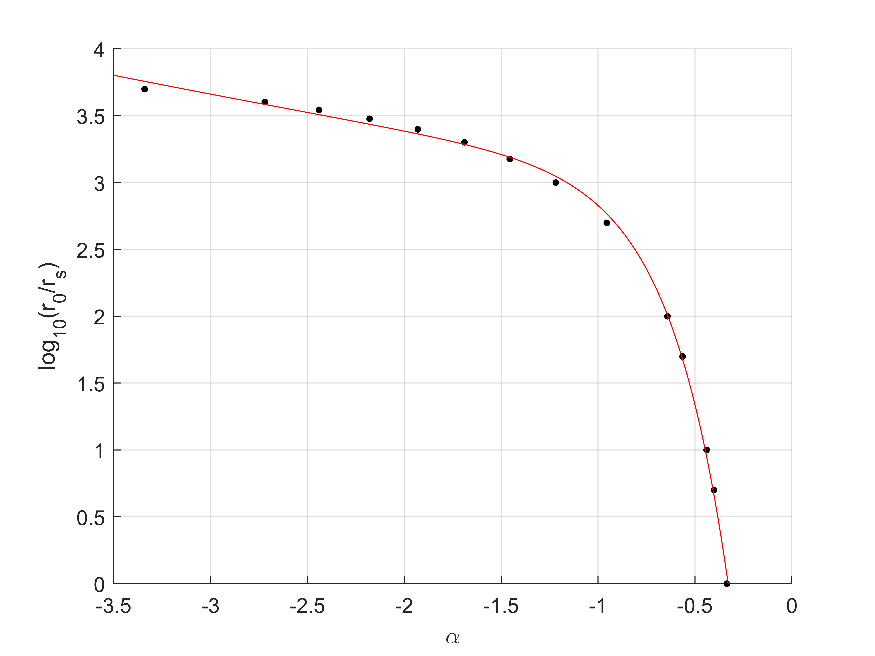}
\includegraphics[trim={0 0.cm 0 0}, clip,width=0.4\textwidth]{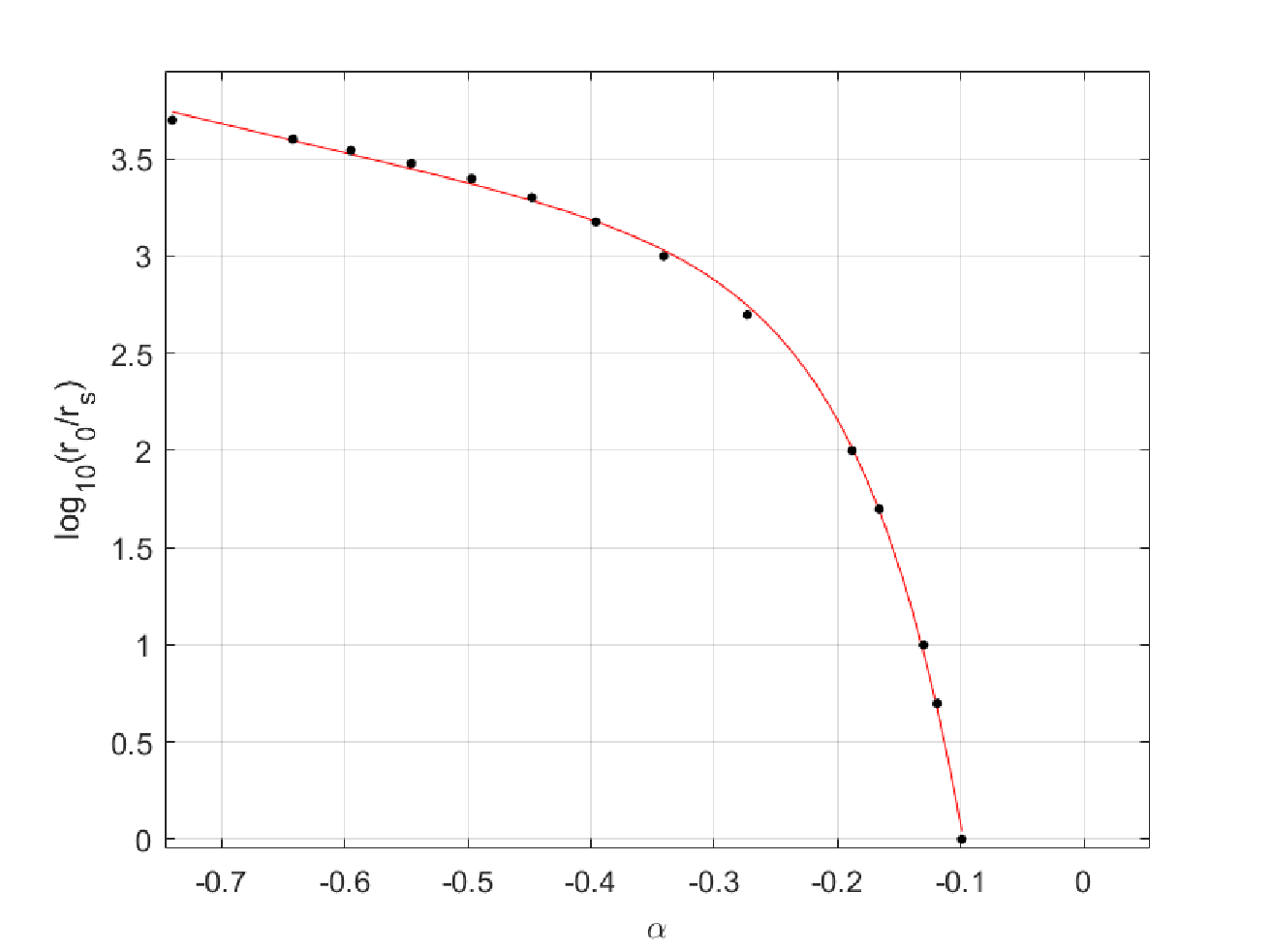}
    \caption{ Dependence of $r_0$ on $\alpha$ for given values of $\dot{M}$. The lines show binomial exponential fits to the data. \emph{Left}: For $\dot{M}=4\times10^{17}$ g s$^{-1}$. The fitting function is $\log_{10}(r_0/r_S)=-15.74e^{2.009\alpha}+3.179e^{-0.04702\alpha}$.  \emph{Right}: For $\dot{M}=4\times10^{18}$ g s$^{-1}$. The fitting function is $\log_{10}(r_0/r_S)=-9.284e^{11.92\alpha}+2.782e^{-0.4012\alpha}$.}
    \label{fig:3}
\end{figure}

{\bf For $r>R_0=3\times 10^4 \ r_S$, since the magnetic pressure is always higher than the gas pressure and the potential energy density, ionized plasmas will be frozen in the large-scale magnetic field in the absence of dissipation, and there will be no accretion.} 
In the following, we will ignore processes for $r>R_0=3\times 10^4 \ r_S$, then the total accretion rate between $r_0$ and $R_0$ is given by:
\begin{equation}\label{eq:Mdot}
\dot{M}=\int_{r_0}^{R_0}4\pi r \rho vdr={4\pi\rho_0 c_s(R_0^{2+\alpha}-r_0^{2+\alpha})\over{r_0^\alpha (2+\alpha)}},
\end{equation}
where $R_0$, $r_0$ are the outer and inner boundaries of the convection dominated accretion flow.
For a given value of the accretion rate $\dot{M}$, the parameters $r_0$ and $\alpha$ are correlated. Figure \ref{fig:3} illustrates this correlation for $\dot{M}=4\times10^{17}$ g s$^{-1}$ (left) and $\dot{M}=4\times10^{18}$ g s$^{-1}$ (right).

\section{Solutions for Accretion Flows and Discussions}

\begin{figure}[h]
    \centering
\includegraphics[trim={0 0.cm 0 0}, clip,width=0.4\textwidth]{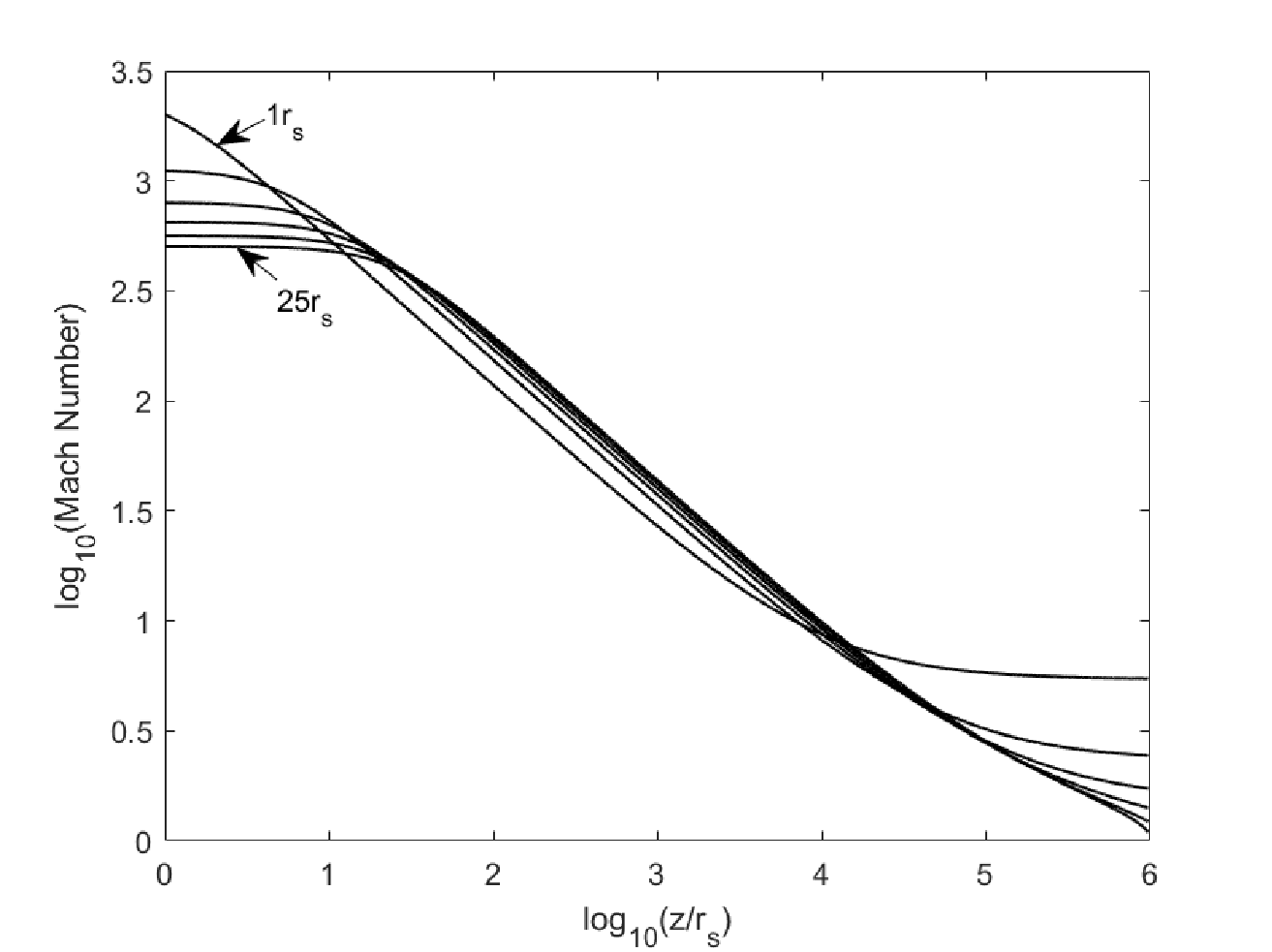}
\includegraphics[trim={0 0.cm 0 0}, clip,width=0.4\textwidth]{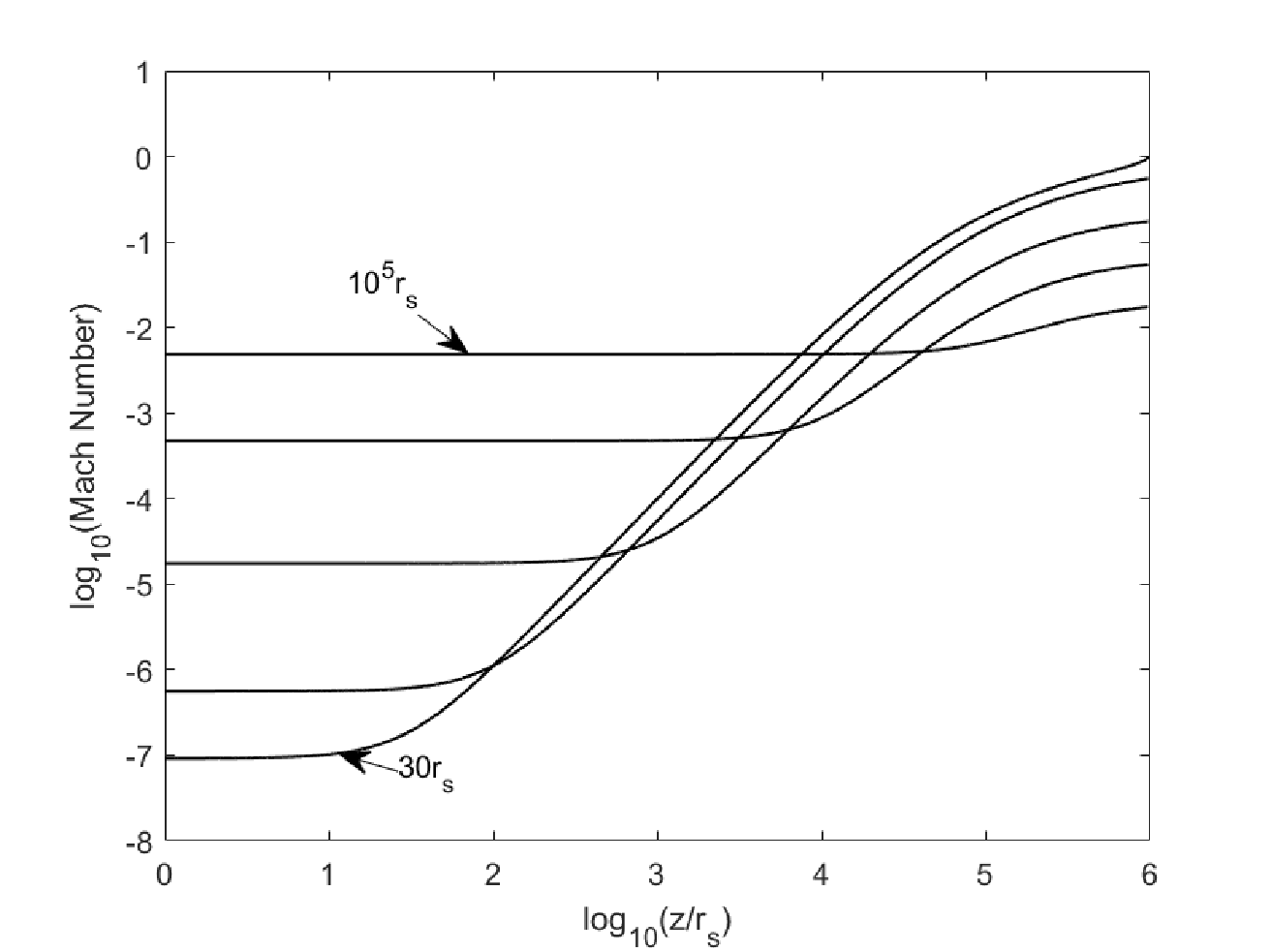}
\includegraphics[trim={0 0.cm 0 0}, clip,width=0.4\textwidth]{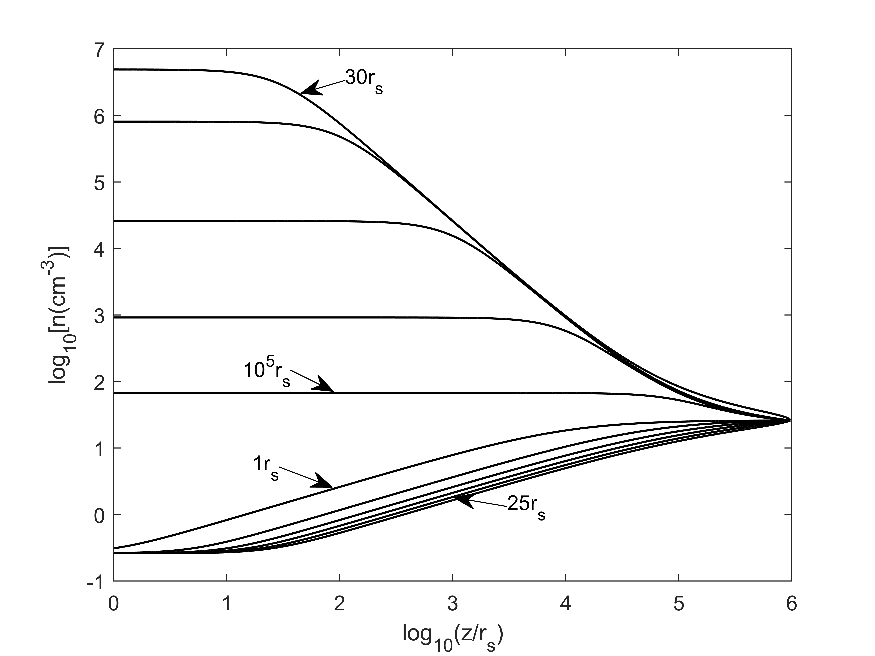}
\includegraphics[trim={0 0.cm 0 0}, clip,width=0.4\textwidth]{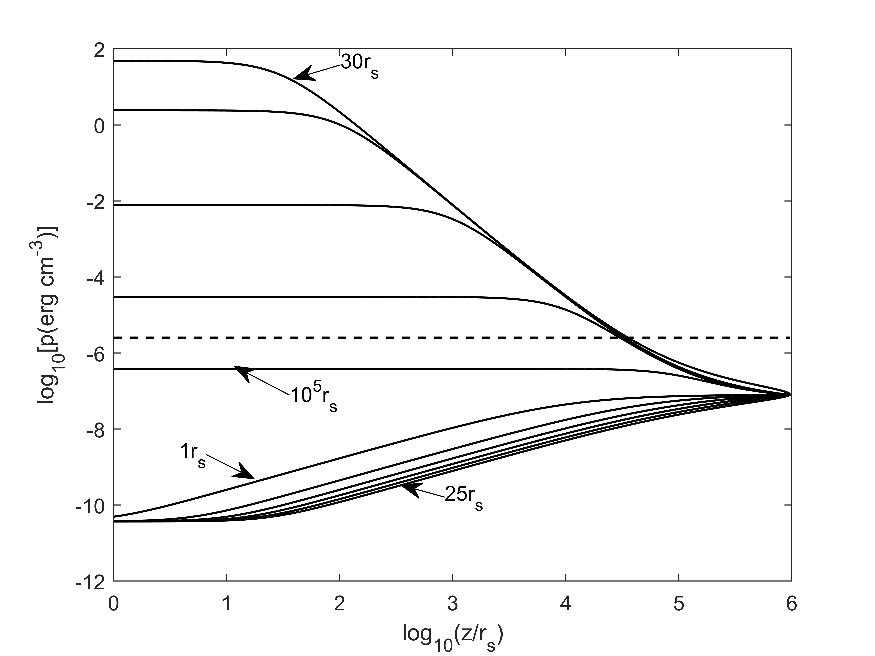}
    \caption{Mach number (upper panels), gas density (lower left) and gas pressure (lower right) profiles with $\rho_{0}=4.3\times 10^{-23}$ g cm$^{-3}$ and $kT = 3.5$ keV at $R=0.4$ pc, $r_0=30\ r_S$, $\alpha=-0.5$, $\gamma={5/3}$. 
    \emph{Top left}: Mach number for $r=1,5,10,15,20,25\ r_S$. \emph{Top right}: Mach number for $r=30,10^2,10^3,10^4,10^5\ r_S$. \emph{Bottom left}: gas density for $r=1,5,10,15,20,25,30,10^2,10^3,10^4,10^5\ r_S$.  \emph{Bottom right}: gas pressure for $r=1,5,10,15,20,25,30,10^2,10^3,10^4,10^5\ r_S$. Solid lines are the gas pressure, and the dashed line is the magnetic pressure for $B=8$ mG.}
    \label{fig:4}
\end{figure}

For given $r_0$, $\alpha$, $\gamma$ and the outer boundary conditions, one can solve equations (1), (2) and (4) to obtain the accretion flow structure along different magnetic field lines. In the following, we set the accretion rate $\dot{M}$ to $4.6\times10^{17}$ g s$^{-1}$ as inferred from modeling of the emission spectrum of Sgr A* \citep{2007ApJ...668L.127L}, 
and adopt the following parameters: $\rho_{0}=4.3\times 10^{-23}$ g cm$^{-3}$ and $kT = 3.5$ keV at $R=0.4$ pc, $r_0=30\ r_S$, $\alpha=-0.5$, $\gamma={5/3}$.
In the one dimensional case, flow along each magnetic field line has two solutions: one is supersonic, and the other subsonic. The outer boundary condition can be used to find the physical solution. 
Equation (4) shows that the flow is supersonic for $r<r_0$ and subsonic for $r\ge r_0$. Figure \ref{fig:4} shows the Mach number, gas density, and pressure profiles for different magnetic field lines. The similarity between the density and pressure profiles within $r_0$ implies a nearly isothermal gas outflow.
{\bf For $r>r_0$ and $z<r$, the gravity does not change significantly with $z$, the density and pressure also do not vary significantly with $z$. For $z>r$, $\rho\sim z^{-3/2}$ and $p\sim z^{5/2}$, similar to the spherically symmetric supersonic accretion flow \citep{1952MNRAS.112..195B}. 
One should note that the solution is invalid in the gas pressure dominated regime, i.e. with $R<R_0=30000r_S$ where one needs to consider the effects of gravity and magnetic energy dissipation, which are needed for the flow to cross field lines and eventually fall into the black hole \citep{2002ApJ...566..137I}. The solution here just gives qualitative estimate of properties of the accretion flow.}

\begin{figure}[h]
    \centering
\includegraphics[trim={0 0.cm 0 0}, clip,width=0.5\textwidth]{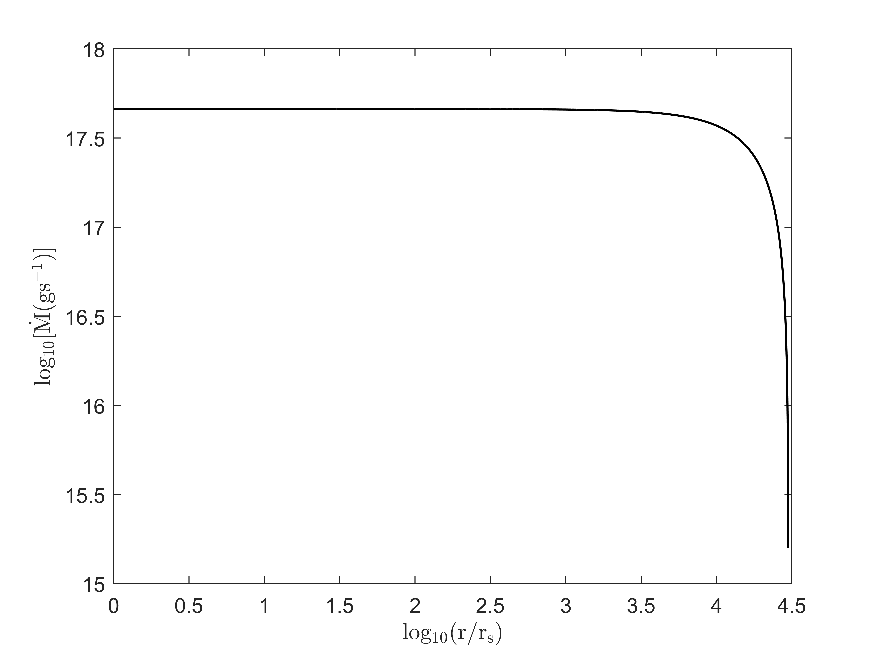}
    \caption{ $\dot{M}$ as a function of $r$ .}
    \label{fig:5}
\end{figure}

As shown in Figure \ref{fig:4}, for $r<r_0$, the Mach number is always greater than 1, indicating a supersonic outflow emerging presumably from corona of the inner accretion flow.  For $r>r_0$, the Mach number is always less than one, indicating a subsonic accretion flow. 
Moreover, it is evident for $r<r_0$, the gas pressure is always less than the magnetic pressure, which explains why the line in Figure \ref{fig:2} is truncated at $r=r_0$ .
Therefore the accretion rate $\dot{M}$ increases from $1.0\ r_S$ to $r_0$, then decreases from r$_0$ to $R_0$, reaching its maximum at $r=r_0$. Figure \ref{fig:5} shows the $r$ dependence of $\dot{M}$.

In 1999, Blandford and Begelman proposed the adiabatic inflow-outflow solution (ADIOS) based on the advection-dominated accretion flows (ADAF) model \citep{1999MNRAS.303L...1B}. ADIOS indicates that ADAF must be accompanied by strong winds, which carry away mass, angular momentum, and energy from the accretion flow.
In the ADIOS, the mass inflow rate satisfies
\begin{equation}\label{massinflowrate}
\dot{m}\propto r^p\ ;\ \ \ \ {\rm with}\ \ \ \     0\leq p <1\, ,
\end{equation}
which implies that the accretion rate increases with radius $r$ following a power law, which is different from our accretion model. 
For $r<r_0$, the accretion model of \citet{2007ApJ...668L.127L} may be applicable and its corona can be connected to the low pressure outflow solution shown here.

\begin{figure}[h]
    \centering
    \includegraphics[trim={0 0.cm 0 0}, clip,width=0.4\textwidth]{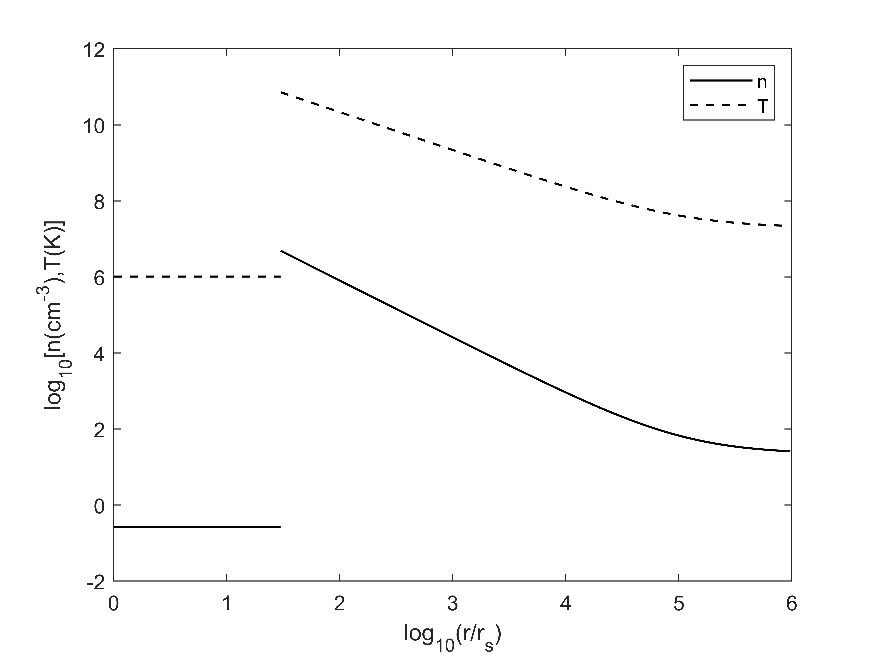}
\includegraphics[trim={0 0.cm 0 0}, clip,width=0.4\textwidth]{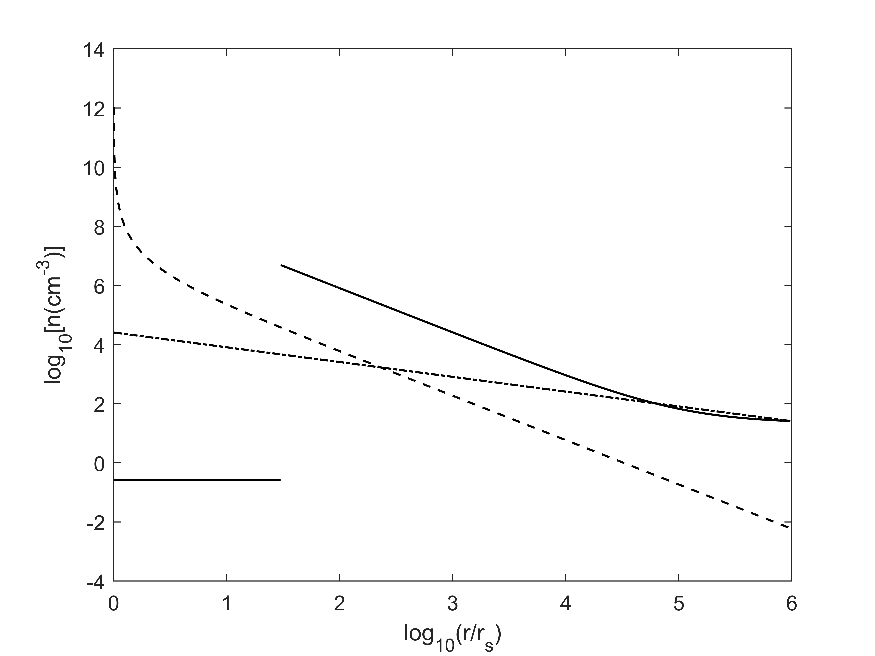}
    \caption{ \emph{Left}:  Gas density (\emph{solid line}) and temperature (\emph{dashed line}) profiles at the equatorial plane \textbf{(equation (1) with z=0 )}. \emph{Right}: Gas density profiles in the equatorial plane (\emph{solid line}) compared with that of a Keplerian two-temperature accretion flow \textbf{[equation (2) in \citet{2007ApJ...668L.127L}]} (\emph{dashed line}).  The $\rho \propto r^{-{1\over2}}$ of the CDAF (\emph{dot-dashed line}) is also shown \citep{2000ApJ...539..798N}. Note that, at small radii, the solution should be interpreted as an outflow from the corona of an accretion disk in the equatorial plane.}
    \label{fig:6}
\end{figure}

Although the flow should be convection dominated between $r_0$ and $R_0$, it is still worth showing properties of gases at the equatorial plane of the one dimensional accretion model. The left panel of Figure \ref{fig:6} shows the density and temperature profiles. {\bf The discontinuity at $r_0$ is caused by using different solutions of the one dimension accretion flow as required by equation (4).}
Since we adopt an ideal adiabatic gas without considering the magnetic field and relativistic effects, the outflow solution within $r_0$ needs to be revised significantly.  \citet{2015ApJ...804..101Y} employed a trajectory method in their numerical simulations of outflows of hot accretion flow, revealing two distinct outflows  above a hot accretion flow: disk jet and wind. The disk jet region is located near the polar axis, characterized by high velocity and low mass flux, primarily driven by magnetic pressure gradients, and the wind has a broader coverage angle, lower velocity, and greater mass flux, primarily driven by the centrifugal force and magnetic pressure gradient. 
Numerical simulations of outflows confirm the presence of strong winds and disk jet in a hot accretion flow.

Above $r_0$, {\bf the density decreases with the increase of $r$ following $\sim r^{-3/2}$ and} is about two orders of magnitude higher than the two temperature Keplerian accretion model proposed by \citet{2007ApJ...668L.127L}. It flattens to $26$ cm$^{-3}$ at the outer boundary of $10^6 r_S\simeq 0.4$pc. Most of the emission is produced within tens of $r_{S}$ in this Keplerian disk model. The differences between the density profiles (right panel of Fig. \ref{fig:6}) can be remedied by considering the convection effect. The CDAF predicts a very shallow density profile (dotted line in the right panel of Fig. \ref{fig:6}) that can match the Keplerian accretion model near $r_0$ \citep{2003ApJ...592.1042I}.
{\bf One should note that although there is no compelling evidence for a Keplerian flow in Sgr A* from EHT observations \citep{2024ApJ...964L..25E}, the model nevertheless reproduces the millimeter spectrum and is used here to constrain properties of the emitting plasma near the black hole \citep{2024ApJ...964L..26E}.}

\section{Discussions and Conclusions}

{\bf The magnetic field of $8$ mG used in this paper is a low limit to explain the observed RM of PSR J1745-2900. A higher value of the magnetic field will lead to a smaller capture radius where the gas pressure starts to dominate. Since the gas pressure scales as $R^{5/2}$, the capture radius should be around $10^4 r_S$. Our model does not consider the effects of stellar winds \citep{2018MNRAS.478.3544R}. The ram pressure of stellar winds can be higher than the magnetic pressure near 0.1 pc from Sgr A* \citep{2020MNRAS.492.3272R}. Disturbance caused by stellar winds may explain the observation variation of RM from  PSR J1745-2900 \citep{2018ApJ...852L..12D}.

In our model, we have not considered the effects of accretion flows on the magnetic field. Assuming that the magnetic field is in energy equipartition with the background plasma and $T\propto R^{-1}$, according to our model $B$ scales as $R^{-5/4}$, RM scales as $R^{-7/4}$. Even considering the convection effects with the density scales as $R^{-1/2}$, $B$ scales as $R^{-3/4}$, the RM scales as $R^{-1/4}$.  Millimeter Observations show that the magnetic field near the black hole is about 100 G \citep{2024ApJ...964L..26E}, implying a magnetic field scaling of $\sim R^{-1}$ from $r_S$ to $10^4 r_S$. Then the gas density needs to scale as $R^{-1}$, which is consistent with MHD simulations \citep{2003ApJ...592.1042I, 2018MNRAS.478.3544R}. In all cases above, mostly of the RM observed from Sgr A* origins from small scale accretion flow, which is consistent with millimeter observations \citep{2024ApJ...964L..25E, 2024A&A...682A..97W, 2022ApJ...930L..19W, 2007ApJ...654L..57M}.}

\textbf{In the work of physical interpretation of the polarized ring by EHT collaboration 
\citep{2024ApJ...964L..26E}, it was demonstrated that when the RM originates from internal Faraday rotation, no model can simultaneously satisfy all total intensity and polarization constraints. With an external Faraday screen, MAD with $a_*=0.94$, $i=150^{\circ}$ and an ion to electron temperature ratio of $160$ is the only viable model with an accretion rate $\approx\ 5\times10^{-9}$ M$_{\odot}$ yr$^{-1}$, several orders of magnitude lower than the Bondi accretion rate. And the model produces a significant outflow power of $4\times10^{38}$ erg s$^{-1}$.
The polarization observations require highly ordered magnetic fields near the event horizon, which is also compatible with MAD.
These ordered magnetic fields may origin from the large scale fields of our model.
}

Considering the effects of a strong magnetic field as revealed with pulsar observations near Sgr A*, we show that accretion starts at a few tens of thousands of Schwarzschild radii of the supermassive black hole in the Galactic center. Assuming that the accretion rate along magnetic field lines at large scales $\rho v$ decreases with the distance $r$ of the magnetic field to the black hole following a power law, we find that the accretion flow model proposed by \citet{2007ApJ...668L.127L} for the sub-millimeter emission from Sgr A* can merge with the CDAF model of \citet{2000ApJ...539..798N} smoothly at a few tens of Schwarzschild radii. 
{\bf Our results point to a self-consistent picture of accretion flows in Sgr A* with the accretion rate strongly suppressed by a strong large scale magnetic field. Detailed numerical simulations with strong large scale magnetic fields can be used to verify this model and explain observations that cannot be addressed with our simple semi-analytical model \citep{2018A&A...618L..10G, 2024ApJ...964L..26E}.}




\begin{acknowledgements}
This work is supported by the National Natural Science Foundation of China under the grants 12375103 and U1931204.
\end{acknowledgements}

\bibliography{myref}
\bibliographystyle{aasjournal.bst}
\end{document}